# Tunable magnetism of a single-carbon vacancy in graphene

Yu Zhang[1,§], Fei Gao[2,§], Shiwu Gao[2,*], and Lin He[1,*]

[1] Center for Advanced Quantum Studies, Department of Physics, Beijing Normal University, Beijing, 100875, People's Republic of China.

[2] Beijing Computational Science Research Center, ZPark II, 100193, Beijing, China.

[§]These authors contributed equally to this work.

[*]Correspondence and requests for materials should be addressed to S.G. (e-mail: swgao@csrc.ac.cn) and L.H. (e-mail: helin@bnu.edu.cn).

**Removing a single-carbon vacancy introduces (quasi-)localized states for both σ and π electrons in graphene. Interactions between the localized σ dangling bond and quasilocalized π electrons of a single-carbon vacancy in graphene are predicted to control its magnetism. However, experimentally confirming this prediction through manipulating the interactions between the σ and π electrons remains an outstanding challenge. Here we report the manipulation of magnetism of individual single-carbon vacancy in graphene by using a scanning tunnelling microscopy (STM) tip. Our spin-polarized STM measurements, complemented by density functional theory calculations, indicate that interactions between the localized σ and quasilocalized π electrons could split the π electrons into two states with opposite spins even when they are well above the Fermi level. Via the STM tip, we successfully manipulate both the magnitude and direction of magnetic moment of the π electrons with respect to that of the σ electrons. Three different magnetic states of the single-carbon vacancy, exhibiting magnetic moments of about 1.6 $\mu_B$, 0.5 $\mu_B$, and 0 $\mu_B$ respectively, are realized in our experiment.**

The miniaturization of fabricating data storage devices with functional building blocks down to the atomic size is a major driving force of nanotechnology, and an outstanding challenge is to control the magnetism at the atomic scale[1,2]. Very recently, the successful introduction of local magnetic moments in graphene by creating atomic vacancies[3] or adsorbing H atoms[4] shows the ability to achieve such a goal. Different from the H adatoms on graphene, which only introduce local magnetic moment of π electrons[4], removing a single-carbon vacancy introduces local magnetic moments of both the π and σ electrons[5-18]. Theoretically, the interactions between the σ dangling bond and the quasilocalized π electrons of the single-carbon vacancy are predicted to control its magnetic states, either exhibiting local magnetic moments or not[6,8,9,11,12,15,16]. Therefore, such a system provides a unique platform to control the magnetism at the atomic scale. In this work, we experimentally studied how the interactions between the σ dangling bond and the quasilocalized π electrons control the magnetism of the single-carbon vacancy in graphene. Via a scanning tunnelling microscopy (STM) tip, we could tune the interactions between the localized σ electrons and the quasilocalized π electron around the single-carbon vacancies and realized three different magnetic states of the single-carbon vacancies in graphene. Density functional theory (DFT) calculations revealed the nature of these magnetic states and quantum phase transitions between the three coupling regimes.

In our experiment, the single-carbon vacancies were generated in the synthesis process of graphene in a chemical vapor deposition (CVD) method (see Methods and Fig. S1 of the Supplemental Material), as reported previously[3,19]. Figure 1a shows a representative atomic STM image of the single-carbon vacancy in graphene, exhibiting the characteristic Jahn-Teller distortion[3,20]. We can clearly observe the triangular $\sqrt{3} \times \sqrt{3}R30°$ interference pattern in the STM image due to the induced intervalley scattering of the atomic defect (see Fig. S2 for more STM images). The single-carbon vacancy exhibits two pronounced peaks in the scanning tunnelling spectroscopy (STS) spectrum (Fig. 1b, Fig. S3 and Fig. S4), attributing to two spin-polarized states of the quasilocalized π electron $V_\pi$ induced by the single-carbon vacancy[3,5].

Figure 1c shows a schematic diagram of electronic structure for a single-carbon vacancy in graphene. The crystal field and Jahn-Teller distortion split the localized σ electrons of the single-carbon vacancy into highly localized $V_{\sigma1}$, $V_{\sigma2}$, and $V_{\sigma3}$ states. Both the $V_{\sigma2}$ and $V_\pi$ states are half occupied because of the electrostatic Coulomb repulsion $U$, i.e., when an electron occupies the state, a second electron with opposite spin must overcome an extra energy $U$. The strength of $U$, which depends on the spatial localization of the states, determines the spin splitting. The spatial extension of the $V_\pi$ state is much larger than that of the $V_{\sigma2}$ state, therefore, the $V_{\sigma2}$ state has a much larger spin splitting than that of the $V_\pi$ state (Fig. 1c). There is usually a single electron occupying the $V_{\sigma2}$ state due to the large spin splitting, however, the two $V_\pi$ states could be double occupied or unoccupied because of the existence of slight charge transfer from the substrate. Then the spin splitting of the $V_\pi$ state should vanish and only one sharp peak is expected to be observed, as reported for the adsorbing H atoms in graphene[4]. However, in our experiment, two peaks are still observed for the single-carbon vacancies even when they are well above the Fermi level, as shown in Fig. 1b and as also reported in a previous work[3]. Our spin-polarized STM measurements, as shown subsequently, demonstrate that the two peaks are the two spin-polarized states of the $V_\pi$ state and we will show that the π magnetism of the single-carbon vacancy behaves quite different from that of the adsorbing H atom in graphene.

The spin-polarized STM measurements, which can directly reflect the spatial distribution of spin-up and spin-down electrons[21,22], provides us an unprecedented insight to identify the π magnetism of the single-carbon vacancy. In our experiment, we used electrochemically etched Ni tips (Fig. S5 and S6 of the Supplemental Material) as the spin-polarized tip. The magnetic polarization of the Ni tip is always along the STM tip due to the weak magnetocrystalline anisotropy of Ni and relatively large shape anisotropy of the STM tip[23,24]. Before the STM measurements, a magnetic field of $B = 3.0$ T ($B = -3.0$ T) perpendicular to the surface of the sample was applied on and then removed gradually to obtain an up-polarized $\vec{M} \uparrow$ (down-polarized $\vec{M} \downarrow$) STM tip.

As schematically shown in Fig. 2a, we can carry out atomic-resolved STS

measurements around the single-carbon vacancy by using spin-polarized STM tip (here the single-carbon vacancy is assumed on a carbon site of the B sublattice). Figure 2b summarizes four representative spin-resolved *dI/dV* spectra measured around the monovacancy (similar STS spectra measured via a nonmagnetic Pt/Ir tip are shown in Fig. S7 of the Supplemental Material for comparison). Because of the spin conservation during the elastic electron tunnelling, the significant differences among these STS spectra directly reflect distributions of spin-polarized density-of-states (DOS) on the A and B sublattices. Our experimental result indicates that the spin-up (spin-down) electrons of the $V_\pi$ state are mainly located on the A (B) sublattice around the monovacancy. Such a result is further confirmed by our STS maps recorded at the energies of the two spin-polarized states (Fig. 2c), revealing the site-specific spin polarization around the single-carbon vacancy, as predicted theoretically[5-18]. The above results explicitly demonstrate that the observed two peaks in the spectra are the two spin-polarized states of the $V_\pi$ state. We attribute the spin splitting of the $V_\pi$ state even when the two spin-polarized states are well above the Fermi level to the interaction between the $V_\pi$ state and the $V_{\sigma 2}$ state. The existence of local magnetic moment from the localized σ electrons is the main difference between the magnetism introduced by the atomic vacancies[3] and by the adsorbing H atoms[4] on graphene. If the graphene is not the exact planar configuration around the single-carbon vacancy, the orthogonality between the σ and π states is broken[6,8,9,11,12,15,16]. Then the coupling with the spin-polarized $V_{\sigma 2}$ state may stabilize the spin splitting of the $V_\pi$ state even when it is double occupied or unoccupied. Therefore, the two states of the $V_\pi$ electrons induced by the atomic vacancies[3] and by the adsorbing H atoms[4] on graphene show quite different behaviors.

The interaction between the localized σ and π states, which depends on the local configuration of graphene, determines the magnetism of the single-carbon vacancy in graphene[6,8,9,11,12,15,16]. Previously, it was demonstrated that the STM tip can generate an out-of-plane displacement on graphene through a finite Van de Waals force[25-28]. Here we show that it is possible to generate out-of-plane lattice deformation around the monovacancy and, consequently, tune interactions between the localized σ and π states by using the STM tip. Figure 3 shows representative results measured on a single-carbon vacancy with three different heights, which are tuned during the scanning process by changing the distance, i.e., the Van de Waals force, between the STM tip and the monovacancy (see Fig. S8 of the Supplemental Material for details). Since both the electronic states and topographic features contribute to the measured height, therefore we measured the relative profile height (apparent height) of the monovacancy under the same experimental condition for comparison. The single-carbon vacancy with different heights exhibits distinct tunnelling spectra, as shown in Fig. 3a-3c. For the case that the measured height of the protrusion around the vacancy is smaller than 50 pm, the energy separation of the two spin splitting states is ~30 meV (Fig. 3a). The relative intensity of the spin-up peak is obvious larger than the spin-down peak when recorded at *A* sublattice, whereas the relative intensity is inversed when recorded at *B* sublattice. Similar features are obtained in tens of the as-grown single-carbon vacancies in graphene by using different STM tips in our experiment. By decreasing the distance between the STM tip and the graphene to about 0.5 nm and scanning the studied monovacancy for about 30 minutes (see Fig S9 and S10 of the Supplemental Material), it is interesting to note that the height of the monovacancy could be increased to above 50 pm. Then the energy difference between the two peaks in the spectrum decreases to ~20 meV and, unexpectedly, the electronic states of the left peak (right peak) are predominantly located on the *B* (*A*) sublattice (Fig. 3b), which are exactly opposite to the result obtained in the as-grown monovacancy. It indicates that the left peak (right peak) of the spectra in Fig. 3b arises from the spin-down (spin-up) state according to the site-specific spin polarization of the quasilocalized π electron. By further decreasing the distance between the STM tip and the graphene to about 0.35 nm, it is possible to tune the height of the monovacancy to above 140 pm (Fig. 3c). In such a case, we only observe a pronounced peak in the tunnelling spectrum. In Fig. 4a, we summarized the energy separations of the two spin-polarized peaks in the STS spectra as a function of the height of the studied monovacancy (for the case that there is only one peak, the energy separation is set as zero). The obtained three distinct spectra indicate that there are three different magnetic states of the monovacancy in graphene, depending on its out-of-plane lattice deformation.

To further understand our experimental result, we carried out DFT calculations on the magnetism of the single-carbon vacancy in graphene as a function of its out-of-plane lattice deformation. Our calculations reveal three different magnetic states of the single-carbon vacancy, which are classified as ferromagnetic (FM), quenched antiferromagnetic (QAFM), and nonmagnetic (NM) phases, as shown in Fig. 4b and 4c. Figures 4d-4f show the corresponding low-energy DOS of the three quantum phases. For the FM phase, the monovacancy exhibits about 1.6 $\mu_B$ local magnetic moments with 1 $\mu_B$ from the localized σ electron and about 0.6 $\mu_B$ from the quasilocalized π electron. The low energy DOS shows two peaks with the spin-up (spin-down) state in the left (right), which is predominantly located on the A (B) sublattice. This agrees quite well with our experimental results obtained in the as-grown monovacancy (Figure 2 and Fig. 3a). Theoretically, the splitting of the two spin-polarized states depends on the size of the graphene super cell. In our calculation, we used a finite size of the graphene super cell due to the limitation of the calculation ability. Therefore, the theoretical splitting is much larger than that observed in the experiment. By increasing the out-of-plane lattice deformation, the monovacancy will become the QAFM phase and its local magnetic moment will decrease to only about 0.5 $\mu_B$, which arises from the fact that the magnetic moments from the localized σ electron and the quasilocalized π electron become antiparallel. In such a phase, the spin splitting is smaller than that of the FM phase by using the same graphene super cell in the calculation. More importantly, the spin-up (spin-down) state, which is predominantly located on the A (B) sublattice, changes to the right (left) peak of the low-energy DOS, as shown in Fig. 4e. These results are well accordant with our experimental results, as shown in Fig. 3a and 3b. By further increasing the out-of-plane lattice deformation of the monovacancy, the strong coupling between the localized σ orbital and π electrons can completely remove the local magnetic moment of the single-carbon vacancy (the NM phase): the local magnetic moments of the localized σ orbital and π orbital exactly cancel with each other. Then, there is only one pronounced peak in the low-energy DOS (Fig. 4f), which is consistent well with our experimental result shown in Fig. 3c. Therefore, it is reasonable to conclude that the observed three distinct spectra of the monovacancy (Fig. 3) correspond to its three different magnetic states.

Our STM measurements, complemented by first-principles calculations, show that the single-carbon vacancy in graphene can exhibit three different magnetic phases depending on the coupling between the localized σ orbital and π electrons. This allows for a consistent interpretation of all current data about the magnetism of the single-carbon vacancy. Previously, distinct tunnelling spectra of single-carbon vacancies in graphene are reported[3,12,14,18]. According to our result, the distinct spectra may arise from the different magnetic phases. The single-carbon vacancies generated in the synthesis process of graphene by the CVD method usually have small apparent height. Therefore, they are mainly in the FM phase and we usually observe two pronounced spin-split peaks in the tunnelling spectra. However, the monovacancy generated by irradiation of high-energy ions are predominantly in the NM phase because of the large apparent height. As a consequence, only a single peak is observed in the spectra[12,14]. In literature[13], the average magnetic moment of the monovacancy generated by irradiation is measured to be about 0.1 $\mu_B$, which is much smaller than the expected valvue 1.6 $\mu_B$, indicating that most of the monovacancies generated by irradiation are in the NM state. In our experiment, we found that about 1/4 of the as-grown monovacancy can be tuned from the FM state to the QAFM state and about 80% of them can be further tuned from the QAFM state to the NM state. However, it is not possible to directly change the defects from FM state into the NM state. Our experiment also demonstrated that both the QAFM and the NM states are quite stable: they cannot relax to the FM state during our experiment and they also cannot be tuned to the FM state by using the STM tip. The substrate may help to stabilize the out-of-plane deformation around the single-carbon vacancy when it was introduced by the STM tip.

In summary, we demonstrated that the magnetism of the single-carbon vacancy is determined by interactions between the localized σ orbital and π electrons. Via an STM tip, we successfully tuned the interactions between the localized σ orbital and π electrons and realized three different magnetic states of the single-carbon vacancy in graphene. Our theoretical calculations further revealed the nature of the three magnetic states. The realizing tunable magnetism at atomic scale may have potential application in graphene-based spintronics.

**Methods**

**Sample preparation of graphene multilayer on Ni foil.**

A traditional ambient pressure chemical vapor deposition (APCVD) method was adopted to grow graphene multilayer on polycrystalline Ni foil. The Ni foil was first heated from room temperature to 1030℃ in 40 min under an argon (Ar) flow of 100 SCCM and hydrogen ($H_2$) flow of 50 SCCM, and keep this temperature and flow ratio for 20 min. Next methane ($CH_4$) gas was introduced with a flow ratio of 20 SCCM to grow graphene for 20 min, and then cooled down to room temperature (see Figure S1).

**STM and STS measurements.** The STM system was an ultrahigh vacuum scanning probe microscope (USM-1300 and USM-1500) from UNISOKU. All STM and STS measurements were performed in the ultrahigh vacuum chamber (~$10^{-11}$ Torr), at liquid-helium temperature (4.2 K) and the images were taken in a constant-current scanning mode. The nonmagnetic STM tips were obtained by chemical etching from a wire of Pt/Ir (80/20%) alloys, and the magnetic STM tips were obtained by electrochemically etching from a wire of Ni. Lateral dimensions observed in the STM images were calibrated using a standard graphene lattice and a Si (111)-(7×7) lattice and Ag (111) surface. The dI/dV measurements were taken with a standard lock-in technique by turning off the feedback circuit and using a 793-Hz 5mV A.C. modulation of the sample voltage.

**Acknowledgements**

This work was supported by the National Natural Science Foundation of China (Grant Nos. 11674029, 11422430, 11374035), the National Basic Research Program of China (Grants Nos. 2014CB920903, 2013CBA01603). L.H. also acknowledges support from the National Program for Support of Top-notch Young Professionals, support from "the Fundamental Research Funds for the Central Universities", and support from "Chang Jiang Scholars Program".

**Author contributions**

Y.Z. synthesized the samples, performed the STM experiments, and analyzed the data. F.G. and S.W.G. performed the theoretical calculations. L.H. conceived and provided advice on the experiment, analysis, and the theoretical calculation. L.H. and Y.Z. wrote the paper. All authors participated in the data discussion.

**Competing financial interests**

The authors declare no competing financial interests.

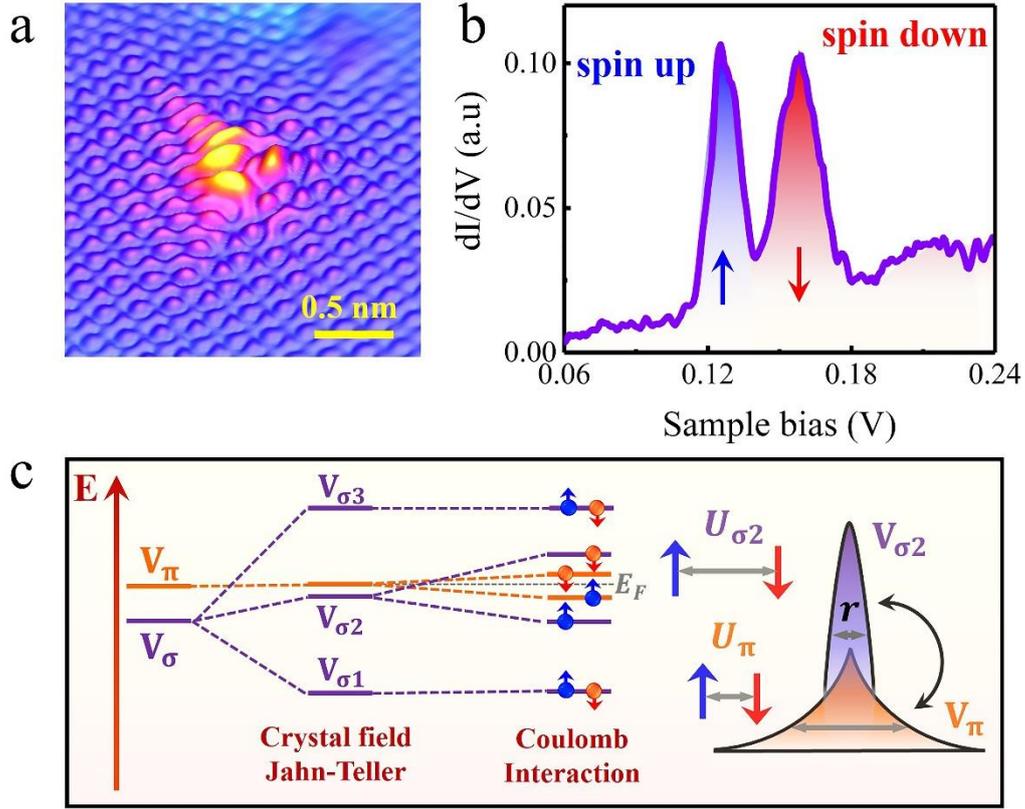

**Figure 1. The spin splitting states induced by a single carbon vacancy in graphene. a,** Atomic resolution STM topography of a single carbon vacancy in the topmost graphene sheet (sample bias: 200 mV, tunneling current: 200 pA). **b,** STS spectrum recorded at the single carbon vacancy. The two peaks reflect the DOS with opposite spin polarizations (sample bias: 200 mV, tunnelling current: 200 pA). **c,** Left panel: Schematic diagram of the electronic structure at a single carbon vacancy in graphene. The removal of a C atom will create three $sp^2\sigma$ dangling bonds adjacent to the new vacancy, and introduce a localized state $V_\sigma$, which further splits into highly localized $V_{\sigma 1}$, $V_{\sigma 2}$, and $V_{\sigma 3}$ states due to the crystal field and Jahn–Teller distortion. Meanwhile, a quasi-localized state $V_\pi$ is also introduced in the band structure. Both of the $V_{\sigma 2}$ and $V_\pi$ states are further splitting because of the electrostatic Coulomb repulsion $U$ at an isolated single carbon vacancy of graphene. Right panel: Illustration of the spin-split states of $V_{\sigma 2}$ and $V_\pi$. The strength of $U$ depends strongly on the spatial localization of the states. The spatial extension of the $V_\pi$ state is much larger than that of the $V_{\sigma 2}$ state, therefore, the $V_{\sigma 2}$ state has a much larger spin splitting than that of the $V_\pi$ state.

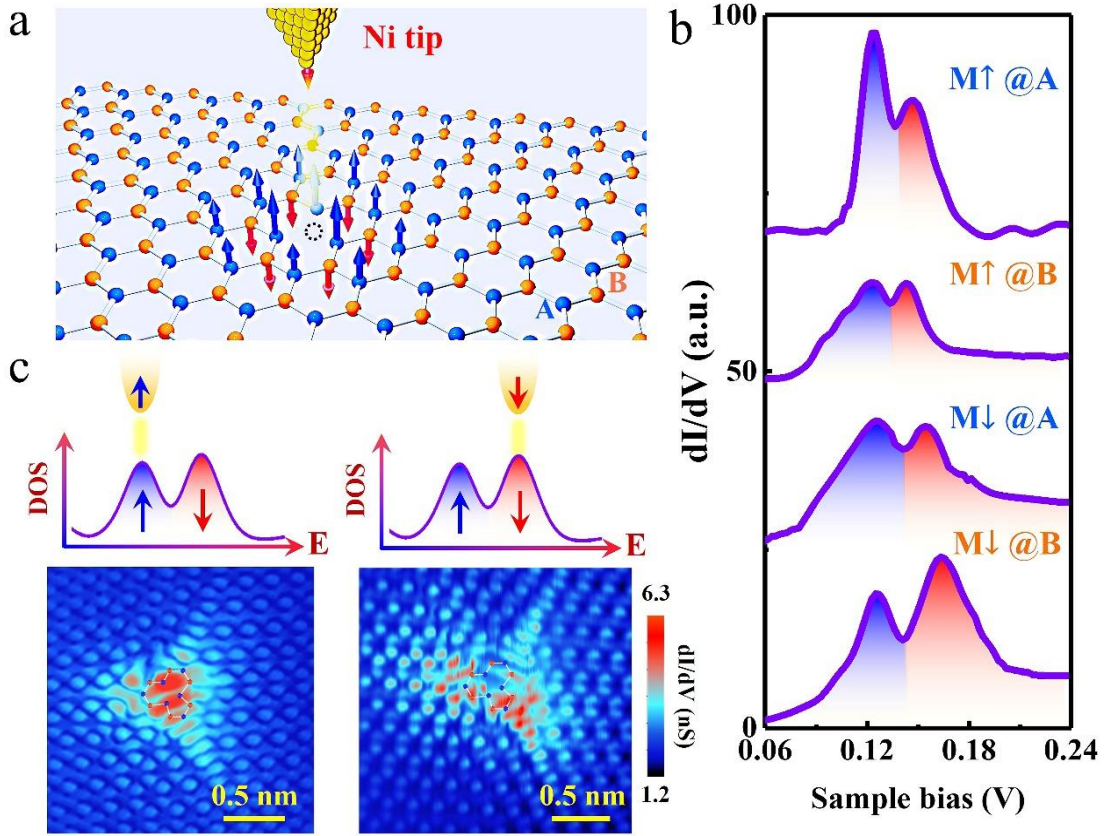

**Figure 2. The spatial distribution of spin-polarized electrons at the single carbon vacancy. a,** Schematic diagram of the spin-polarized STS. The black dashed circle shows the position of the missing C atom, and the arrows reflect the spin directions. The spin is conserved during the elastic electron tunnelling, and the spins of sample aligned perpendicular to the sample surface can be detected by the magnetic Ni tip along the tip axis. **b,** The spin-polarized *dI/dV* spectra of tip polarization $\vec{M}\uparrow$ ($\vec{M}\downarrow$) on A (B) sublattice respectively. **c,** The spin-polarized differential conductance maps at the bias voltages of the spin up (low energy) and spin up (high energy) states respectively.

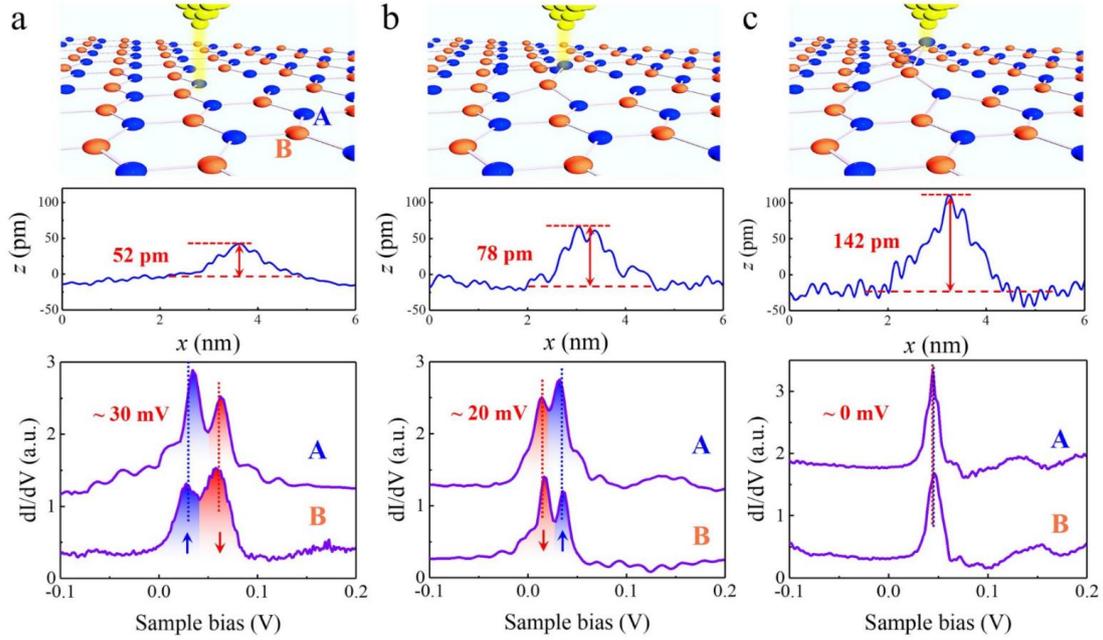

**Figure 3. Tunable magnetism of the single-carbon vacancy in graphene. a-c,** Up panel: Schematic side view of the configuration around the vacancy induced by the van der Waals attraction between the STM tip and the graphene. Middle panel: Typical heights of the single carbon vacancy (sample bias: 500 mV, tunneling current: 150 pA). The heights of the protrusion are ~52 pm, ~78 pm, and ~142 pm respectively. Bottom panel: the corresponding *dI/dV* spectra recorded at *A* and *B* sublattices of the single carbon vacancy. The relative intensities of two localized peaks recorded at *A* and *B* sublattice reflect the distributions of the DOS with opposite spin polarizations.

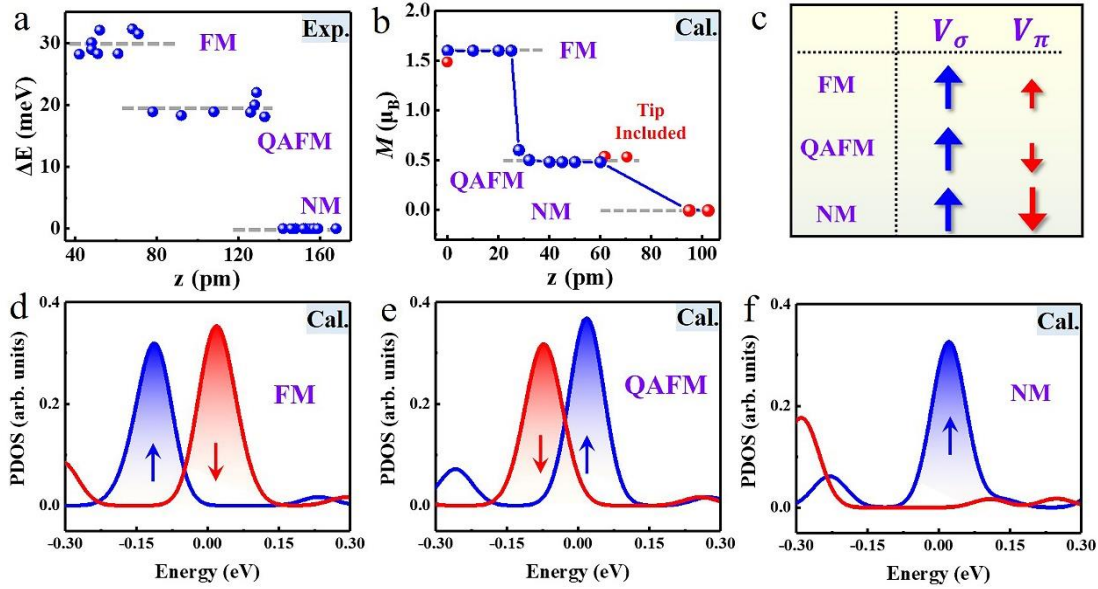

**Figure 4. Quantum phase transition of FM, QAFM, and NM states. a,** The energy separations of the two spin-polarized DOS peaks in STS spectra as a function of the height of the studied monovacancy in STM image. **b,** Theoretical calculations of the localized magnetic moments of the single-carbon vacancy in graphene as a function of its out-of-plane lattice deformation. The localized magnetic moments of $\sim 1.6\,\mu_B$, $\sim 0.5\,\mu_B$, and $\sim 0\,\mu_B$ represent the FM, QAFM and NM states, respectively. The blue dots indicate the intrinsic magnetic moments of the defects with out-of-plane lattice deformation, and the red dots indicate the induced magnetic states by considering the interaction between the $V_\pi$ state and STM tip. **c,** Schematic diagram of the spin alignment of $V_\sigma$ and $V_\pi$ moments with the change of out-of-plane lattice deformation at the single-carbon vacancy. **d-f,** The localized states of graphene with a single carbon vacancy in the $8\times 8$ supercell are obtained from the extensive first-principles calculations, which represent the FM, QAFM, and NM states respectively. The states of spin up and spin down electrons are indicated by blue and red lines.